\documentclass[11pt]{article}

\usepackage{amsmath}
\usepackage{amssymb}
\usepackage{epsfig}
\usepackage{color} 
\label{•} 
\begin{document}

%%%%%%%%%%%%%%%%%%%%%%%%%%%%%%%%%%%%%%%%%%%%%%%%%%%%%%%%%%%%%%%%%%%%%

\begin{titlepage}

% the footnote symbols are only redefined for the title page !
\renewcommand{\thefootnote}{\alph{footnote}}
\vspace*{-3.cm}
\begin{flushright}

\end{flushright}

\vspace*{0.3in}

\renewcommand{\thefootnote}{\fnsymbol{footnote}}
\setcounter{footnote}{-1}

{\begin{center} {\Large\bf  $SO/Sp$ Chern-Simons Gauge Theories At Large $N$,\\  $SO/Sp$ Penner Models\\ And The Gauge Group Volumes}

\end{center}}
\renewcommand{\thefootnote}{\alph{footnote}}

\vspace*{.8cm} {\begin{center} {\large{\sc
                Noureddine~Chair$^{1\dagger}$\footnote{$\dagger$ Author to whom correspondence should be addressed}, Mohammad Dalabeeh$^2$
                }}
\end{center}}
\vspace*{0cm} {\it
\begin{center}
 $^1$Physics Department,
 University of Jordan, Amman, Jordan

Email: n.chair@ju.edu.jo\\ \hspace{19mm}\\ $^2$Physics Department, Jerash University, Jerash, Jordan\\Email: madalabeeh@jpu.edu.jo

\end{center} }

\vspace*{1.5cm}

{\Large \bf
\begin{center} Abstract
\end{center} }
\ 
We construct a deformed $SO/Sp$ Penner generating function responsible for the close connection between $SO/Sp$ Chern-Simons gauge theories at large $N$ and the $SO/Sp$ Penner models. This construction is then shown to follow from a sector of a Chern-Simons gauge theory with coupling constant $\lambda$. The free energy and its continuum limit of the perturbative Chern-Simons gauge theory are  obtained from the  Penner model. Finally, asymptotic expansions for the logarithm of the gauge group volumes  are  given for every genus $g\geq 0$ and shown to be equivalent to the continuum limits of the $SO/Sp$ Chern-Simons gauge theories and the $SO/Sp$ Penner models.
 \vspace*{.5cm}
\end{titlepage}
\renewcommand{\thefootnote}{\arabic{footnote}}
\setcounter{footnote}{0}
\newpage
\section{Introduction}
The free energy of the Penner model \cite{penner,harer} is the generating function of the orbifold Euler characteristic of the moduli space of Riemann surfaces  of genus $g$, with $n$ punctures. The $SU(N)$ perturbative Chern-Simons free energy, based on the $1/N$ expansion introduced by 't Hooft \cite{thooft}, and the Penner free energy have a similar   topological expansion. The perturbative Chern-Simons free energy  \cite{gobkumar} 
may be written as 
%%%%%%%%%%%%%%%%%%%%%%%
\begin{equation}
F=\sum_{g=0, h=1}C_{g, h}N^{2-2g}\lambda^{2g-2+h},
\end{equation}
%%%%%%%%%%%%%%%%%%%%%%%%%%%
where $\lambda$ is the 't Hooft coupling constant and $h$ is the number of faces (boundaries) of the triangulated Riemann surfaces. In the Penner model $h$ is identical to the number of punctures.
The coefficient $C_{g, h}$ was shown by Witten \cite{witten} to be identical to the partition function of the A-model topological open string theory of genus $g$ with $h$ boundaries on a six-dimensional target space $T^*S^3$. It has been shown  by the first author \cite{chair} that the coefficient $C_{g, h}$ are related to the orbifold Euler characteristic of the moduli space of Riemann surfaces of genus $g$ with $2h$ punctures. For  the perturbative $SO(N)$ Chern-Simons gauge theory \cite{sinha}, the topological expansion when $N$ is even has the form 
%%%%%%%%%%%%%%%%%%%%%%%%%%%%%%%%%%%%%%%%%%%
\begin{equation}
F=\sum_{g=0, h=1}C_{g,2h}(N-1)^{2-2g}\lambda^{2g-2+2h}+\sum_{g=0, h=1}\tilde{C}_{g, 2h+1}(N-1)^{1-2g}\lambda^{2g+2h},
\end{equation}
%%%%%%%%%%%%%%%%%%%%%%%%%%%%%%%%%%%%%%%%%%%%%%
where the first term corresponds to half of  the topological expansion  of the $SU(N)$ Chern-Simons free energy, while  the second term is nothing but the topological expansion of the non-orientable $SO(N)$ Chern-Simons  free energy. We will show  that the coefficients  $\tilde{C}_{g,2h+1}$ are related to the orbifold Euler characteristic of the non-orientable Riemann surfaces of genus $g$ with $2h+1$ punctures \cite{l.chekhov,jakson}. Our goal in this work is to push further the   connection between the $SU(N)$ Chern-Simons and the Penner model observed by the first author \cite{chair}, when the gauge group is $SO(N)/Sp(N)$. It has been shown \cite{chair}, that  computations in the perturbative $SU(N)$ Chern-Simons may be carried out using the Penner model. Here, we construct a deformed $SO(N)$ Penner generating function and show that it gives rise to the perturbative $SO(N)$ Chern-Simons gauge theory. Such a construction turns out to contain both the $SO(N)$ Penner generating function \cite{mulase,madalabeeh}, and a generating function for the non-orientable orbifold Euler characteristic with coupling constant $t/2$. A check for such a construction is given by computing the free energy of the perturbative $SO(N)$ Chern-Simons theory on $S^3$ in terms of the string coupling constant $g_s$ and the K\"ahler parameter  $t$, this will be given in section \ref{sec3}. We provide a proof for  our  construction  based on a simple observation that the perturbative $SO(N)$ Chern-Simons gauge theory splits into two sectors, one with coupling constant $\lambda$ and the other sector with coupling constant $-\lambda$, this will be given in section \ref{sec2}. The perturbative free energy ${F}^{SO}(\lambda,N)$ associated with  coupling constant  $\lambda$ is shown to generate the orbifold Euler characteristic of the moduli space of  orientable and non-orientable  Riemann  surfaces of genus $g$ with $n$ punctures. Das and Gomez \cite{das}  reproduced the nonperturbative terms in the $SU(N)$ Chern-Simons theory using the continuum limit of the perturbative $SU(N)$ Chern-Simons theory. The nonperturbative terms are known to be related to the volume of the $SU(N)$ gauge group \cite{oguri},  these terms also have been obtained  using the Penner model  \cite{chair}. In section \ref{continuum}, of this paper, the continuum limit for the perturbative $SO(N)/Sp(N)$ Chern-Simons gauge theories are  obtained and shown to be related to the asymptotic expansions of $\log ( \text{vol}(SO(2N+1)))$ and $\log( \text{vol}(Sp(2N-1)))$ respectively, given in section \ref{nonpert}. The asymptotic expansions of $\log (\text{vol}(G))$, for  $G=SO(2N+1)$, $G=Sp(2N-1)$ are derived explicitly for  all genera $g\geq 0$, this is  shown  to  follow simply from  the asymptotic expansions of the Barnes and the Gamma functions \cite{oguri, Adamchik}. Also,  the asymptotic expansion of $\log( \text{vol}(SO(2N)))$  and  $\log( \text{vol}(Sp(2N)))$ were shown to reproduce the continuum limit of the $SO(2N)/Sp(2N)$ Penner models derived in \cite{madalabeeh}. Finally, in the last section  our work is summarized, and the relations  between the gauge group volumes are deduced as well as the relation between $\log\frac{\text{vol}(SO(2N))}{\text{vol}(Sp(2N))}$ and the generating function for the orbifold Euler characteristic associated with the non-orientable Riemann surfaces of genus $g$ with $n$ punctures.
%%%%%%%%%%%%%%%%%%%%%%%%%%%%%%%%%%  
\section{The $SO(N)/Sp(N)$ Penner Models \label{section 1}}
%%%%%%%%%%%%%%%%%%%%%%%%%%%
 Here, we will  briefly review the  $SO(N)/Sp(N)$ Penner models and its continuum limit studied in  \cite{madalabeeh}. 
The free energy of the $SO(N)/Sp(N)$ Penner models are the generating functions of the 
orbifold Euler characteristic of the moduli space of real  algebraic curves \cite{jakson}. 
The explicit expression for the topological expansion of the  $SO(N)/Sp(N)$   Penner free energy $F^{So/Sp}$ \cite{mulase} in 
terms of the genus $g$ and the punctures $p$ is:
%%%%%%%%%%%%%%%%%%%%%%%%%%%%%%%%%
\begin{equation}
\label{so/sp}
\begin{split}
F^{So/Sp}(t,N)=&\frac{1}{2}\sum_{\substack{g\ge 0, p>0\\2-2g-p<0}}
\chi^O(\mathfrak M_{g,p}) (2N)^p (t)^{2g+p-2}\\
&\mp\sum_{\substack{g\ge 0,p>0\\ 1-2g-p<0}}
 \chi^{NO}(\mathfrak M_{g,p})
(2N)^{p} (t)^{2g+p-1},
\end{split}
\end{equation}
%%%%%%%%%%%%%%%%%%%%%%%%%%%%%%%5
where $\chi^{O}(\mathfrak M_{g,p})$ and $\chi^{NO}(\mathfrak M_{g,p})$
are the orbifold Euler characteristic of the moduli space of complex and real algebraic curves respectively, given by:
%%%%%%%%%%%%%%%%%%%%%%%%%%%%%%%%%%%
 \begin{equation}
\label{chior}
\chi^{O}(\mathfrak M_{g,p})=(-1)^p \frac{(2g+p-3)!(2g-1)}{(2g)!p!}B_{2g}, \nonumber
\end{equation}     
\begin{equation}
\label{chinor}
\chi^{NO}(\mathfrak M_{g,p})=(-1)^p\frac{1}{2}\frac{(2g+p-2)!(2^{2g-1}-1)}{(2g)!p!}B_{2g},
\end{equation}
%%%%%%%%%%%%%%%%%%%%%%%%%%%%%%%%%%%%%%%%%%%%%
where $B_{2g}$ are the Bernoulli numbers.  The first term in the $SO(N)/Sp(N)$   Penner free energy $F^{So/Sp}$ corresponds to the 
orientable surfaces contributions, and is equal
to  half of the free energy of the ordinary  Penner model 
 with the size of the matrix doubled, while the second term 
corresponds to  the  non-orientable   contributions. 
The  partition  function of $SO(N)/Sp(N)$ Penner models   $ Z(t,N)^{So/Sp}=e^{F(t,N)^{So/Sp}}$,  may be written as  \cite{madalabeeh}: 
%%%%%%%%%%%%%%%%%%%%%%%%%%%%%%%%%%%%%5
\begin{equation}
\label{penner symplectic}
Z(t,N)^{So/Sp}=\Bigr[(\frac{\sqrt{2\pi t}(et)^{-(t^{-1})}}{\Gamma(\frac{1}{t})})^{N}\prod_{p=1}^{2N}(1+pt)^{\frac{1}{2}(2N-p)}\Bigl]\prod_{p\ \text{odd}}^{2N-1}(1+pt)^{\pm\frac{1}{2}}.
\end{equation}
%%%%%%%%%%%%%%%%%%%%%%%%%%%%%%%%%%5
It is clear from this expression that $Z(t,N)^{So/Sp}$ is the product of two   partition functions  in which  the first is ordinary Penner model \cite{penner,chairold},  given 
by the term between the square brackets, while   the second is the  partition function for the   non-orientable  contributions coming from $SO(N)$ / $Sp(N)$  Penner models 
\cite{madalabeeh}. By making the natural scaling limit $t\rightarrow \frac{-t}{2N}$ in the free energy $F^{SO/Sp}(t,N)$,
the continuum limit of the  $SO(N)/Sp(N)$ Penner models \cite{madalabeeh} may be  obtained either  using the Euler-Maclaurin formula or by
summing over all punctures in the expression for the free energy given by equation (\ref{so/sp}). Then, one defines a new coupling constant
 $\mu=2N(1-t)$, such that $\mu$ is kept fixed as  $N\rightarrow\infty$, and $t\rightarrow 1$ (the double scaling limit). Having done so, the expression for the free energy in the continuum limit becomes 
%%%%%%%%%%%%%%%%%%%%%%%%%%%%%%%%%%
\begin{equation}
\label{finalsym}
F(\mu)^{So/Sp}=\frac{1}{4}\mu^2 \log\mu\mp\frac{1}{4}\mu\log\mu
-\frac{1}{24}\log\mu\pm\frac{1}{24\mu}+\frac{1}{2}\sum_{g\geq2}\frac{1}{(2g-2)}\frac{B_{2g}}{2g}\mu^{2-2g}\pm\sum_{g\geq 2}^{\infty}\frac{(2^{2g-1}-1)}{(2g-1)}\frac{B_{2g}}{4g}\mu^{1-2g},
\end{equation}
%%%%%%%%%%%%%%%%%%%%%%%%%%%%%%5
where the coefficients of $\mu^{2-2g}$ and $\mu^{1-2g}$ are the topological orbifold Euler characteristic without punctures $\chi^{O}(\mathfrak M_{g,0})$ and $ \chi^{NO}(\mathfrak M_{g,0})$ respectively. It is interesting to note that the $SO/Sp$ Penner models and the Penner model share the same critical point $t=1$ \cite{madalabeeh}.
%%%%%%%%%%%%%%%%%%%%%%%%%%%%%%%%%%%%%%
\section{Perturbative $SO(N)$ Chern-Simons Theory From The Orthogonal
Penner Model}\label{sec3}
%%%%%%%%%%%%%%%%%%%%%%%%%%%%%%%%%%%%%%
In this section we propose a deformed $SO(N)$ Penner generating function that gives rise to the perturbative $SO(N)$ Chern-Simons gauge theory. 
This is motivated by the  established connection between the ordinary Penner model and the  $SU(N)$ Chern-Simons gauge theory given by the first 
author \cite{chair}. However, this time our proposed generating function  will 
contain both the $SO(N)$ Penner model as well as a term that generates the non-orientable orbifold Euler characteristic with coupling 
constant $\frac{t}{2}$\footnote{We set the size of the matrix   
equal to $2N$ for both the orthogonal and the symplectic Penner models,
while in \cite{sinha} the size of the matrix   is set to be $N+a$, where $a=-1$ for
the orthogonal case and $a=-1$ for the symplectic case, where they  become identical
in the  large $N$ limit.}.
To that end, let us consider a combination of the orthogonal Penner models defined as follows:
%%%%%%%%%%%%%%%%%%%%%%%%%%%%%%%
\begin{equation} 
\label{math construction}
\mathcal{F}(t,N)=F^{SO}(t,N)-2F^{SO/Non}(\frac{t}{2},N), 
\end{equation} 
%%%%%%%%%%%%%%%%%%%%%%%%%%%%%%%%%%%%
where $F^{SO/Non}(\frac{t}{2},N)$  corresponds to the  generating function for the non-orientable orthogonal Penner model with
coupling constant $\frac{t}{2}$.
The above deformed orthogonal Penner model in the scaling $t \rightarrow
\frac{t}{2N}$, may be rewritten as
%%%%%%%%%%%%%%%%%%%%%%%%%%%%%%%%%% 
\begin{eqnarray}
\label{extop}
\mathcal{F}(t,N)&=&\frac{1}{2}\sum_{\substack{g\ge 0, p>0\\2-2g-p<0}} 
 \chi^{O}(\mathfrak M_{g,p})(2N)^{2-2g} (t)^{2g+p-2}\nonumber \\ 
&-&\sum_{\substack{g\ge 0,p>0\\ 1-2g-n<0}}
 \chi^{NO}(\mathfrak M_{g,p})
(2N)^{1-2g} (t)^{2g+p-1}\nonumber  \\ &+&2\sum_{\substack{g\ge 0,p>0\\ 1-2g-p<0}}
 \chi^{NO}(\mathfrak M_{g,p})
(2N)^{1-2g} (\frac{t}{2})^{2g+p-1}.
\end{eqnarray}
%%%%%%%%%%%%%%%%%%%%%%%%%%%%%%%%%%%%%%%
Let us now  consider a sum of two deformed orthogonal Penner models, 
one with coupling constant $t$, and the other with coupling 
constant $-t$, such that the topological expansion of the free 
energy in both cases is given by equation (\ref{extop}). If the coupling constant in the deformed orthogonal  
Penner model is set to be equal to $\lambda/2\pi n$, where $\lambda$ is the Chern-Simons coupling constant and $n$
is a positive integer, and let $\mathbf F(\lambda,N)$ be the total free energy for the two deformed $SO(N)$ Penner models summed over $n$,
then by using equation (\ref{extop}) one has
%%%%%%%%%%%%%%%%%%%%%%%%%%%%%%%%%%%%
\begin{eqnarray}
\label{seven}
\mathbf{F}(\lambda,N)&=&\frac{1}{2}\sum_{\substack{g\ge 0, p>0\\2-2g-p<0}}\sum_{n=1}^\infty 
 \chi^{O}(\mathfrak M_{g,p})(2N)^{2-2g} (\lambda/2\pi n)^{2g+p-2}(1+(-1)^p)\nonumber \\
 &+&2\sum_{\substack{g\ge 0,p>0\\ 1-2g-p<0}}\sum_{n=1}^\infty
 \chi^{NO}(\mathfrak M_{g,p})(2N)^{1-2g} (\lambda/2\pi n)^{2g+p-1}(\frac{1}{2^{2g+p-1}}-\frac{1}{2})(1+(-1)^{p-1}). 
\end{eqnarray}
%%%%%%%%%%%%%%%%%%%%%%%%%%%%%%%%%%%%%
The first term on  the right hand side of the above equation (\ref{seven}), contributes to the total free energy only when  $p$  is even,
while   the second term  contributes only when $p$ is  odd. Therefore, the total free energy  $\mathbf{F}(\lambda,N)$ becomes 
%%%%%%%%%%%%%%%%%%%%%%%%%%%%%%%%
\begin{eqnarray}
\label{connn}
\mathbf{F}(\lambda,N)&=&\sum_{\substack{g\ge 0, p>0\\1-g-p<0}} 
 \sum_{n=1}^\infty\chi^{O}(\mathfrak M_{g,2p})(2N)^{2-2g} (\lambda/2\pi n)^{2g+2p-2}\nonumber \\
 &+&4\sum_{\substack{g\ge 0, p>0\\ g+p>0}}
 \sum_{n=1}^\infty\chi^{NO}(\mathfrak M_{g,2p+1})(2N)^{1-2g} (\lambda/2\pi n)^{2g+2p}(\frac{1}{2^{2g+2p}}-\frac{1}{2}).
\end{eqnarray}
%%%%%%%%%%%%%%%%%%%%%%%%%%%%%%%%%%%% 
This result may be compared with the work of Sinha  and Vafa \cite{sinha},  in which the first and the  second  terms in  equation (\ref{connn})  are  nothing but the orientable and the  non-orientable  surfaces  contributions  to the free energy of the  $SO(N)$ Chern-Simons gauge  theory respectively.  At the level of the partition function, the connection between the perturbative $SO(N)$  Chern-Simons and the $SO(N)$ Penner partition functions may be written as  
\begin{equation}
{Z}^{SO}_{CS}(\lambda)=\prod_{n=1}^{\infty}{Z}_d^{SO}(\lambda){Z}_d^{SO}(-\lambda),
\end{equation}
where ${Z}_d^{SO}(\lambda)$ stands for the partition function of the deformed $SO(N)$ Penner model with  coupling constant $\lambda$,   given by 
\begin{equation}
{Z}_d^{SO}(\lambda)=[(\frac{\sqrt{\lambda/n}(e\lambda/2\pi n)^{\frac{-2\pi n}{\lambda}}}{\Gamma(2\pi n/\lambda)})^N\prod_{p=1}^{2N}(1+p\lambda/2\pi n)^{\frac{1}{2}(2N-p)}]\prod_{p\ \text{odd}}^{2N-1}\frac{(1+p\lambda/2\pi n)^{\frac{1}{2}}}{(1+p\lambda/4\pi n)}.
\end{equation}
Now, using this partition function ${Z}_d^{SO}(\lambda)$, and making the natural scaling $\lambda\rightarrow \frac{\lambda}{2N}$ then  one can show that\footnote{The term $\frac{\sqrt{\lambda/n}(e\lambda/2\pi n)^{\frac{-2\pi n}{\lambda}}}{\Gamma(2\pi n/\lambda)}$ may be expanded as in  \cite{chairold, mulase}. However this term will be canceled out upon multiplying  ${Z}_d^{SO}(\lambda)$ by ${Z}_d^{SO}(-\lambda)$.} 
\begin{equation}
{Z}^{SO}_{CS}(\lambda)=\prod_{n=1}^{\infty}[\prod_{p=1}^{2N}(1-(p\lambda/4N\pi n)^2)^{\frac{1}{2}(2N-p)}]\prod_{p\ \text{odd}}^{2N-1}\frac{(1-(p\lambda/4N\pi n)^2)^{\frac{1}{2}}}{(1-(p\lambda/8N\pi n)^2)}.
\end{equation}
The first term between the square brackets corresponds to the partition function of the $SU(N)$ Chern-Simons gauge theory, while, the second term  may be  identified with the non-orientable  contribution to the partition function of the perturbative  $SO(N)$ Chern-Simons gauge theory.  Explicitly, the $g=0$,  free energy reads
\begin{eqnarray}
%%%%%%%%%%%%%%%%%%%%%%%%%%%%%%
\mathbf{{F}}^0(\lambda,N)&=&\sum_{n=1}^{\infty}\sum_{p=1}^\infty 
 \frac{-1}{2p(2p-1)(2p-2)}(2N)^{2} (\frac{\lambda}{2\pi n})^{2p-2}\nonumber \\ 
&+&\sum_{n=1}^{\infty}\sum_{p=1}^\infty
 \frac{1}{(2p+1)(2p)}(2N) (\frac{\lambda}{2\pi n})^{2p}(\frac{1}{2^{2p}}-
 \frac{1}{2}).\nonumber \\
\end{eqnarray}
%%%%%%%%%%%%%%%%%%%%%%%%%%%%%%%%
Similarly for genus $g\geq 1$ one has\footnote{To see this we make the following shift  $p\rightarrow p-1$ in equation (\ref{connn}).}
%%%%%%%%%%%%%%%%%%%%%%%%%%%%%%%%%%5
\begin{eqnarray}
\mathbf{{F}}(\lambda,N)&=&\sum_{n=1}^{\infty}\sum_{\substack{g\ge 1, p>0\\1-g-p<0}} 
 \frac{(2g+2p-3)!(2g-1)}{(2g)!(2p)!}B_{2g}(2N)^{2-2g} (\frac{\lambda}{2\pi n})^{2g+2p-2}\nonumber \\ 
&-&2\sum_{n=1}^{\infty}\sum_{\substack{g\ge 0,p>0\\ g+p>0}}
 \frac{(2g+2p-1)!(2^{2g-1}-1)}{(2g)!(2p+1)!}B_{2g}(2N)^{1-2g} (\frac{\lambda}{2\pi n})^{2g+2p}(\frac{1}{2^{2g+2p}}-
 \frac{1}{2}).\nonumber \\
\end{eqnarray}
%%%%%%%%%%%%%%%%%%%%%%%%%%%%%%%%%%%5}}}
Therefore,  using the  deformed $SO(N)$ Penner model described  by equation  (\ref{math construction}), we succeeded in obtaining  the perturbative $SO(N)$ Chern-Simons theory given by equation (\ref{connn}). The latter  was obtained originally  using the large $N$ expansion for the 
 partition function of the Chern-Simons theory for $SO(N)$ and $Sp(N)$ gauge groups \cite{sinha}.
Now the  computation in the  Chern-Simons gauge theory may be carried out  using the
 deformed  orthogonal Penner model and this will be done in the next subsections.
 %%%%%%%%%%%%%%%%%%%%%%%%%%%%%%%%
\subsection{ $SO(N)$ Chern-Simons Gauge Theory At Large $N$ From  The $SO(N)$ Penner Model
}\label{gzero}
%%%%%%%%%%%%%%%%%%%%%%%%%%%%%%%%%%
%%%%%%%%%%%%%%%%%%%%%%%%%%%%%%%%%%%%%%%
In  this section we are going to use the deformed orthogonal
Penner generating function to compute the  worldsheet $RP2$ contribution, as well as the higher genus contribution to the sum over all punctures of the  free energy of the $SO(N)$ Chern-Simons gauge theory \cite{sinha}. This serves as a mere check of our proposal.
As  the deformed $SO(N)$ Penner model contains half of the ordinary Penner model  \cite{madalabeeh}, and  the fact that  the $SO(N)$ Chern-Simons theory  already contains   half the $SU(N)$ Chern-Simons theory \cite{sinha}. Then, we   will concentrate only on the non-orientable contributions,  since  computations for the orientable contributions may be found in  \cite{chair}. 
%%%%%%%%%%%%%%%%%%%%%%%%%%%%%%%%%%%%%%
\subsubsection{The Genus $g=0$  Computation}
%%%%%%%%%%%%%%%%%%%%%%%%%%%% 
The non-orientable contribution for the free energy from equation (\ref{extop}), when $g=0$   reads    
%%%%%%%%%%%%%%%%%%%%%%%%%%%%%%%5
\begin{eqnarray}
\label{11}
\mathcal{F}^{NO}_{0}(t,N)=-\frac{N}{2}\sum_{p=2}
 \frac{1}{p(p-1)}
(-t)^{p-1}+N\sum_{p=2}
 \frac{1}{p(p-1)}
(\frac{-t}{2})^{p-1},
\end{eqnarray}
%%%%%%%%%%%%%%%%%%%%%%%%%%%%%%%%%%%%
in obtaining the above equation, we have used the expression for $\chi^{NO}(\mathfrak M_{0,p})$  given by equation (\ref{chinor}).
The sum  over boundaries (punctures) may be carried out  using the  identity 
%%%%%%%%%%%%%%%%%%%%%%%%%%%%%%%%%%% 
\begin{equation}
\sum_{p=2}^{\infty}
 \frac{1}{p(p-1)}
({-t})^{p-1}=\Bigr{[}1-(\frac{1+t}{t})\log(1+t)\Bigl{]},
\end{equation}
%%%%%%%%%%%%%%%%%%%%%%%%%%%%%%%%%%%%%%
therefore, the genus zero contribution becomes 
%%%%%%%%%%%%%%%%%%%%%%%%%%%%%%%%%%
\begin{eqnarray}
\label{contg=0}
\mathcal{F}^{NO}_{0}(t,N)=\frac{2N}{t}\Bigr{[}\frac{t}{4}-({1+t/2})
\log(1+t/2)+\frac{1}{4}({1+t})\log(1+t)\Bigl{]}.
\end{eqnarray}
%%%%%%%%%%%%%%%%%%%%%%%%%%%%%%%%%%%%
Using the established  relation between the perturbative $SO(N)$ Chern-Simons  and the deformed $SO(N)$ Penner model summarized by equation (\ref{seven}), the sum over boundaries for the total free energy  $\mathbf{F}^{NO}_{0}(\lambda,N)$, is 
%%%%%%%%%%%%%%%%%%%%%%%%%%%%%%%%%%%%
\begin{eqnarray}
\mathbf{F}^{NO}_{0}(\lambda,N)=\frac{2N}{\lambda}\sum_{n=1}^{\infty}\Bigr{[}\frac{\lambda}{2}+2\pi n({1-\lambda/4\pi n})
\log(1-\lambda/4\pi n)-\frac{\pi n}{2}({1-\lambda/2\pi n})\log(1-\lambda/2\pi n)\nonumber \\-2\pi n({1+\lambda/4\pi n})
\log(1+\lambda/4\pi n)+\frac{\pi n}{2}({1+\lambda/2\pi n})\log(1+\lambda/2\pi n)\Bigl{]},
\end{eqnarray}
%%%%%%%%%%%%%%%%%%%%%%%%%%%%%%%%%%
or in a more compact form
%%%%%%%%%%%%%%%%%%%%%%%%%%%%%%%%%%%%
\begin{eqnarray}
\label{z}
\mathbf{F}^{NO}_{0}(\lambda,N)=\frac{2N}{\lambda}\sum_{{n\in {\bf Z}, n\neq 0}}\Bigr{[}\frac{\lambda}{4}+2\pi n({1-\lambda/4\pi n})\log(1-\lambda/4\pi n)\nonumber \\ -\frac{\pi n}{2}({1-\lambda/2\pi n})\log(1-\lambda/2\pi n)\Bigl{]}.
\end{eqnarray}
%%%%%%%%%%%%%%%%%%%%%%%%%%%%%%%%%%%%
The free energy may be written in terms of the worldsheet instanons $\exp(-t)$ as in \cite{sinha}, to do so we 
let  $\mathbf{F}^{NO}_{0}(\lambda,N)=(\frac{2N}{\lambda})\mathbf{F}^{NO}_{0}(\lambda)$   then  differentiate   $\mathbf{F}^{NO}_{0}(\lambda)$  with respect to $\lambda$,  to give
%%%%%%%%%%%%%%%%%%%%%%%%%%%%%%%%%%%
\begin{equation}
\label{eta}
\frac{d}{d\lambda}{\mathbf{F}^{NO}_{0}(\lambda)}=\sum_{{n\in {\bf Z}, n\neq 0}}\Bigr{[}\frac{1}{4}\log(1-\lambda/2\pi n)-\frac{1}{2}\log(1-\lambda/4\pi n)\Bigl{]}.
\end{equation}
%%%%%%%%%%%%%%%%%%%%%%%%%%%%%%%%%%
The sum in the above equation may be carried out  using the following identity \footnote{This identity  is obtained  from the  product formula
 $\frac{sin\pi x}{\pi x}=\prod_{n=1}^{\infty}(1-\frac{x^2}{n^2})$.}
 %%%%%%%%%%%%%%%%%%%%%%%%%%%%%%%%%%%%%%%%%% 
\begin{eqnarray}
\sum_{{n\in {\bf Z}, n\neq 0}}\log(1-\lambda/{2\pi n})=\frac{i\lambda}{2}
+\log(1-e^{-i\lambda})-\log{\lambda}-\frac{i\pi}{2},
\end{eqnarray}
%%%%%%%%%%%%%%%%%%%%%%%%%%%%%%%%%%%%%%
to obtain  
%%%%%%%%%%%%%%%%%%%%%%%%%%%%%%%%%%%%%%%%%%
\begin{eqnarray}
\frac{d}{d\lambda}{\mathbf{F}^{NO}_{0}(\lambda)}=\bigr{[}\frac{1}{4}
\log{\frac{1+e^{-i\lambda/2}}{1-e^{-i\lambda/2}}}+\frac{1}{4}\log{\lambda}+\frac{i\pi}{8}-\frac{1}{2}\log2\bigl
{]}.
\end{eqnarray}
%%%%%%%%%%%%%%%%%%%%%%%%%%%%%%%%%%%%%%%%%%%
From the identity
%%%%%%%%%%%%%%%%%%%%%%%%%%%%%%%%%%%%%
\begin{eqnarray}
\label{exp}
\log{\frac{1+e^{-i\lambda/2}}{1-e^{-i\lambda/2}}}=2\sum_{n\ \text{odd}}^{\infty}\frac{e^{-in\lambda/2}}{n},
\end{eqnarray}
%%%%%%%%%%%%%%%%%%%%%%%%%%%%
 the differentiated free energy  $\frac{d}{d\lambda}{\mathbf{F}^{NO}_{0}(\lambda)}$,   reads 
 %%%%%%%%%%%%%%%%%%%%%%%%%%%5
\begin{eqnarray}
\label{etadef}
\frac{d}{d\lambda}{\mathbf{F}^{NO}_{0}(\lambda)}=\frac{1}{2}\sum_{n\ \text{odd}}^{\infty}\frac{e^{-in\lambda/2}}{n}+\frac{1}{4}\log{\lambda}+\frac{i\pi}{8}-\frac{1}{2}\log2.
\end{eqnarray}
%%%%%%%%%%%%%%%%%%%%%%%%%%%%%%%%%%%%%%%
Now integrating   $\frac{d}{d\xi}{\mathbf{F}^{NO}_{0}(\xi)}$ with respect to $\xi$ from $0$ to $\lambda$, and   using the substitution $\lambda=-it$,   where $t$ is now is the K\"ahler parameter and replace $\frac{2N}{\lambda}$ by $\frac{i}{g_s}$ where  $g_s$ is  the string coupling constant,  then  the total free energy $\mathbf F_0^{NO}(t,N)$,  reads
 %%%%%%%%%%%%%%%%%%%%%%%%%%%%%%%%%%%
\begin{equation} 
\label{fo}
\mathbf{F}^{NO}_{0}(t,N)\cong\frac{1}{g_s}
\sum_{n\ \text{odd}}\frac{e^{-nt/2}}{n^2},
\end{equation} 
%%%%%%%%%%%%%%%%%%%%%%%%%%%%%%%%%%%%
this expression is in complete 
agreement with that 
 obtained  by  Sinha and  Vafa \cite{sinha},  using computations in 
the perturbative $SO(N)$ Chern-Simons theory.
%%%%%%%%%%%%%%%%%%%%%%%%%%%%%%%%%%%%%%%%%%%
\subsubsection{Higher Genus Computations}\label{highergenus}
%%%%%%%%%%%%%%%%%%%%%%%%%%%%%%%%%%%%%%%%%%%%
By following the same procedure used  for the $g=0$,
the higher genus contributions  to the  free energy of $SO(N)$ Chern-Simons theory in terms of the worldsheet instantons may be obtained using the deformed $SO(N)$ Penner model. From equation (\ref{extop}), the 
non-orientable surfaces contributions to the deformed $SO(N)$  Penner model is 
%%%%%%%%%%%%%%%%%%%%%%%%%%%%%%%%%%%%%
%%%%%%%%%%%%%%%%%%%%%%%%%%%%%%%%%%%%% 
\begin{eqnarray}
\label{highergenuss}
\mathcal{F}^{NO}(t,N)&=&\sum_{\substack{g\ge 1,p>0\\ 1-2g-p<0}}
 (-1)^p\frac{(2g+p-2)!(2^{2g-1}-1)}{(2g)!\;p!}B_{2g}
(2N)^{1-2g} (\frac{t}{2})^{2g+p-1}\nonumber \\ &-&\frac{1}{2}\sum_{\substack{g\ge 1,p>0\\ 1-2g-p<0}}
 (-1)^p\frac{(2g+p-2)!(2^{2g-1}-1)}{(2g)!\;p!}B_{2g}
(2N)^{1-2g} (t)^{2g+p-1}.
\end{eqnarray}
%%%%%%%%%%%%%%%%%%%%%%%%%%%%%%%%%%%%%%
%%%%%%%%%%%%%%%%%%%%%%%%%%%%%%%%%%%
From  the identity
 %%%%%%%%%%%%%%%%%%%%%%%%%%%%%%%%%%%%%
 %%%%%%%%%%%%%%%%%%%%%%%%%%%
\begin{equation} 
\frac{d^p}{dt^p}(1+t)^{1-2g}=(-1)^p\frac{(2g+p-2)!}{(2g-2)!}(1+t)^{1-2g-p},
\end{equation}
%%%%%%%%%%%%%%%%%%%%%%%%%%%%%%%%%%%%%%%%
%%%%%%%%%%%%%%%%%%%%%%%%%%%%%%%%%%%%%%%%
together with Maclaurin series expansion of the function $(1+t)^{1-2g}$, 
the summation over punctures in  equation (\ref{highergenuss}) reads 
%%%%%%%%%%%%%%%%%%%%%%%%%%%%%%%%%%%%%%%%%
%%%%%%%%%%%%%%%%%%%%%%%%%%%%%%%%%%%%%%%%%5
\begin{eqnarray}
\label{higher genus}
\mathcal{F}^{NO}(t,N)=\sum_{g\ge 1}\frac{2^{2g-1}-1}{4g(2g-1)}B_{2g}
\biggr{[}t^{2g-1}-2(\frac{t}{2})^{2g-1}+\big{(}\frac{2N(1+t)}{t}\bigl{)}^{1-2g}-2\big{(}\frac{2N(1+t/2)}{t/2}\bigl{)}^{1-2g}\biggl{]}.
\end{eqnarray}
%%%%%%%%%%%%%%%%%%%%%%%%%%%%%%%%%%%%%%%%
Again  using the established  relation between the perturbative $SO(N)$ Chern-Simons  and the deformed $SO(N)$ Penner model, the sum over boundaries for the total free energy is 
%%%%%%%%%%%%%%%%%%%%%%%%%%%%%%%%%%%%%%%%%
\begin{eqnarray}
\label{lambdasum}
\mathbf{F}^{NO}(\lambda,N)=\sum_{g\ge 1}\frac{2^{2g-1}-1}{4g(2g-1)}B_{2g}
(\frac{\lambda}{2N})^{2g-1} \sum_{\substack{n\in \mathbb{Z} \\ n\neq 0}}
\biggr{[}2\big{(}\frac{1}{4\pi n+{\lambda}}\bigl{)}^{2g-1}-\big{(}\frac{1}{2\pi n+{\lambda}}
\bigl{)}^{2g-1}\biggl{]}.
\end{eqnarray}
%%%%%%%%%%%%%%%%%%%%%%%%%%%%%%%%%%
The expression for the total free energy $\mathbf{F}^{NO}(\lambda,N)$ given by the above relation may be rewritten using the 
 following  relations\footnote{Note that the terms that are not written in the approximation $\sum_{\substack{n\in \mathbb{Z} \\ n\neq 0}}\log(1+\frac{\lambda}{2\pi n})\cong \log(1-e^{-i\lambda})$ would disappear upon differentiation $(2g-1)$  times.}
\begin{equation*}
\label{lsum} 
 \frac{d^{2g-1}}{d\lambda^{2g-1}}\log(1+\frac{\lambda}{2\pi n})=\frac{(2g-2)!}{(2\pi n+{\lambda})^{2g-1}},
\end{equation*} 
\begin{equation*}
\sum_{\substack{n\in \mathbb{Z} \\ n\neq 0}}\log(1+\frac{\lambda}{2\pi n})\cong \log(1-e^{-i\lambda}),
\end{equation*}
%%%%%%%%%%%%%%%%%%%%%%%%%%%%%%%%%%%%%%%%%%%
together with  the identity given by equation (\ref{exp}), as follows 
%%%%%%%%%%%%%%%%%%%%%%%%%%%%%%%%%
\begin{eqnarray}
\label{lambdasum5}
\mathbf{F}^{NO}(\lambda,N)\cong - \sum_{g\ge 1}\frac{2^{2g-1}-1}{(2g)!}B_{2g}(\frac{\lambda}{2N})^{2g-1} 
\frac{d^{2g-1}}{d\lambda^{2g-1}}\sum_{p\ \text{odd}}^{\infty}\frac{e^{-in\lambda/2}}{n}.
\end{eqnarray}
%%%%%%%%%%%%%%%%%%%%%%%%
Carrying out the differentiation, we obtain 
%%%%%%%%%%%%%%%%%%%%%%%%%%%%%%
\begin{eqnarray}
\label{lambdasum6}
\mathbf{F}^{NO}(\lambda,N) \cong \sum_{g\ge 1}\frac{B_{2g}}{(2g)!}(\frac{\lambda}{2N})^{2g-1}
\bigr{(}1-\frac{1}{2^{2g-1}}\bigl{)}(i)^{2g-1}\sum_{p\ \text{odd}}^{\infty}n^{2g-2}{e^{-in\lambda/2}}.
\end{eqnarray}
%%%%%%%%%%%%%%%%%%%%%%%%%%%%%%%
In terms of  the  string variable $g_s$, and K\"ahler parameter $t$, the above expression becomes
%%%%%%%%%%%%%%%%%%%%%%%%%%%%%%%
\begin{eqnarray}
\label{t7}
\mathbf{F}^{NO}(g_s,t)\cong \sum_{g\ge 1}\frac{B_{2g}}{(2g)!}g_{s}^{2g-1}
\bigr{(}1-\frac{1}{2^{2g-1}}\bigl{)}\sum_{p\ \text{odd}}^{\infty}n^{2g-2}{e^{-nt/2}}.
\end{eqnarray}
%%%%%%%%%%%%%%%%%%%%%%%%%%%%%%
Equivalently, this may be rewritten in terms of the orbifold Euler characteristic without punctures $\chi^{NO}(\mathfrak M_{g,0})$,  associated with moduli space of non-orientable surfaces as 
%%%%%%%%%%%%%%%%%%%%%%%%%%%%%
\begin{eqnarray}
\label{t70}
\mathbf{F}^{NO}(g_s,t)\cong \sum_{g\ge 1}\frac{\chi^{NO}(\mathfrak M_{g,0})}{(2g-2)!}g_{s}^{2g-1}
\sum_{p\ \text{odd}}^{\infty}(\frac{n}{2})^{2g-2}{e^{-nt/2}}.
\end{eqnarray}
%%%%%%%%%%%%%%%%%%%%%%%%
As it is clear from this expression there are no constant maps - those maps for which the whole Riemann surface of genus $g$ is mapped to a point. This can be understood from the deformed $SO(N)$ Penner free energy given by equation (\ref{higher genus}).
Now, the coefficient of the first and the second terms will disappear when constructing the   perturbative  Chern-Simons free energy, since both of these terms are odd in $t$.  This is unlike the  $SU(N)$ case where the term related to the the constant maps is even in $t$ \cite{chair}, and when constructing the perturpative $SU(N)$ Chern-Simons free energy their coefficient is identified with the Hodge integral. 
%%%%%%%%%%%%%%%%%%%%%%%
Now going back to equation (\ref{t7}), and taking into account  contributions  from $g=0$  and using the following identity \cite{identity} 
%%%%%%%%%%%%%%%%%%%%%%%%%%%%%%%%%%%%%%%%%%%%
\begin{equation*}
\text{csch}(\frac{ng_s}{2})=\sum_{g=0}\frac{2B_{2g}}{(2g)!}\bigr(1-2^{2g-1}\bigl)\Bigr(\frac{ng_s}{2}\Bigl)^{2g-1},
\end{equation*}
%%%%%%%%%%%%%%%%%%%%%%%%%%%%%%%%%%%%%
Then the full free energy takes the form
%%%%%%%%%%%%%%%%%%%%%%%%%%%%%%%%%
\begin{eqnarray}
\label{t8}
\mathbf{F}^{NO}(g_s,t)\cong -\sum_{p\ \text{odd}}\frac{e^{-nt/2}}{2n\sinh(ng_s/2)}.
\end{eqnarray}
%%%%%%%%%%%%%%%%%%%%%%%%%%%%%%%%%%%%%
This is  in a  complete agreement with  \cite{Vincent} (see equation (3.2)). The restriction on the sum over  $n$ may be lifted to give 
%%%%%%%%%%%%%%%%%%%%%%%%%%%%%%%%%%%%%%%%%%
\begin{eqnarray}
\label{t9}
\mathbf{F}^{NO}(g_s,t)\cong - \frac{1}{2}\Bigr[\sum_{n=1}\frac{e^{-nt/2}}{2n\sinh(ng_s/2)}-
\sum_{n=1}(-1)^n\frac{e^{-nt/2}}{2n\sinh(ng_s/2)}\Bigl].
\end{eqnarray}
%%%%%%%%%%%%%%%%%%%%%%%%%%%%%%%%%%%%%%%%%%
Now, we may extend our work to include the $Sp(N)$ case  as well. To do so,  we note  that the  $SO(N)/SP(N)$ free energies  of the  Penner models differ by a minus sign in front of the non-orientable part contribution this is the duality between $SO(N)/Sp(N)$ Penner models \cite{mulase}\footnote{More precisely $F^{SO}(t,N)=F^{Sp}(-t,-N)$.}. Also this duality   is  present in the $SO(N)/Sp(N)$ perturbative Chern-Simons gauge theories. Therefore, it  follows  that   defining a deformed $Sp(N)$ Penner model as in equation (\ref{math construction}) will give rise to the perturbative $Sp(N)$ Chern-Simons. Hence, 
 %%%%%%%%%%%%%%%%%%%%%%%%%%%%%%%%%5
\begin{eqnarray}
\label{t10}
\mathbf{F}_{Sp}^{NO}(g_s,t)\cong \frac{1}{2}\Bigr[\sum_{n=1}\frac{e^{-nt/2}}{2n\sinh(ng_s/2)}-
\sum_{n=1}(-1)^n\frac{e^{-nt/2}}{2n\sinh(ng_s/2)}\Bigl],
\end{eqnarray}
%%%%%%%%%%%%%%%%%%%%%%%%%%%%%%%%%%%%%%%%%
this  is exactly equivalent to set $t$ equals to $t+2\pi i$ in equation (\ref{t9}) as noted in  \cite{sinha},
that is,
%%%%%%%%%%%%%%%%%%%%%%%%%%%%%%%%%%%%%%
\begin{equation}
\mathbf{F}_{Sp}^{NO}(g_s,t)=\mathbf{F}_{SO}^{NO}(g_s,t+2\pi i).
\end{equation}
%%%%%%%%%%%%%%%%%%%%%%%%%%%%%
\section{From The Perturbative $SO(N)$ Chern-Simons Gauge Theory  To The $SO(N)$ Penner Model}\label{sec2}
%%%%%%%%%%%%%%%%%%%%%%%%%%%%%%%%%%%%%%%%%%%
In section \ref{sec3}, we obtained the perturbative $SO(N)/Sp(N)$ Chern-Simons gauge theories  using the $SO(N)/Sp(N)$ Penner models by construction. Here, we will see that the former  generates the orbifold Euler characteristic of the moduli space of orientable and non-orientable  Riemann surfaces of genus $g$ with $n$ punctures, this in turns  proves our construction. Therefore  the Penner model  may be thought of as a building block for the Chern-Simons gauge theory.
The free energy  of the perturbative $SO(N)/Sp(N)$ Chern-Simons gauge theories   is given by the following term \cite{sinha} 
%%%%%%%%%%%%%%%%%%%%%%%%%%%%%%%%%%%%%%5
\begin{eqnarray}
\label{new1}
\mathbf{F}^{SO/Sp}&=&\log Z(\lambda, N) \nonumber \\
&=&\sum_{j\geq 1}^{N-2}f(j)\Bigr{[}\sum_{p=1}^{\infty}\log\bigr{(}1-\frac{j^2\lambda^2}{4\pi^2p^2(N+a)^2}\bigl{)}\Bigl{]},
\end{eqnarray}
%%%%%%%%%%%%%%%%%%%%%%%%%%%%%%%
for $SO(N)$ $a=-1$ and if $N$ is even then the weight $f(j)$ is given by \cite{sinha}
%%%%%%%%%%%%%%%%%%%%%%%%%%%%%%%%%%%%%
\begin{equation}
f(j) =
\begin{cases}
\frac{(N+1-j)}{2} & j\ \text{odd}\  j < N/2,\\
\frac{(N-1-j)}{2} & j\ \text{odd}\  j \geq N/2,\\
\frac{(N-j)}{2} & j\ \text{even}\  j < N/2,\\
\frac{(N-j-2)}{2} & j\ \text{even}\  j \geq N/2.
\end{cases}
\end{equation}
%%%%%%%%%%%%%%%%%%%%%%%%%%%%%%%%%%
It is a simple   observation   that the free energy in equation (\ref{new1}) factorizes into two sectors  one with coupling constant $\lambda$, and the other with coupling constant  $-\lambda$ as 
%%%%%%%%%%%%%%%%%%%%%%%%%
\begin{equation}
\mathbf{F}^{SO}=F^{SO}(\lambda,N)+F^{SO}(-\lambda,N).
\end{equation}
%%%%%%%%%%%%%%%%%%%%%%%%%%%%%%%%%%%%
Although, this is a  simple observation it has a deep consequences, to see this let us consider the first part of the free energy in the above equation \footnote{The following computations are similar to those of Sinha and Vafa \cite{sinha}, see section four therein, however, here we will concentrate on one sector of the perturbative Cheren-simons free energy.},  that is,  
%%%%%%%%%%%%%%%%%%%%%%%%%%%%
\begin{eqnarray}
F^{SO}(\lambda,N)&=&\sum_{j\geq 1}^{N-2}f(j)\Bigr{[}\sum_{p=1}^{\infty}\log\bigr{(}1+\frac{j\lambda}{2\pi p(N-1)}\bigl{)}\Bigl{]} \nonumber \\
&=&- \sum_{j\geq 1}^{N-2}f(j)\sum_{m=1}^{\infty}j^m\Bigr(\frac{-\lambda}{2(N-1)\pi}\Bigr{)}^m\frac{\zeta(m)}{m},
\end{eqnarray}
the sum over $j$ gives
\begin{equation}
\sum_{j\geq 1}^{N-2}f(j)j^m=\sum_{j\geq 1}^{N-2}\frac{(N-1-j)}{2}j^m+\sum_{j\geq 1}^{N/2-1}j^m-2^{m-1}\sum_{j\geq 1}^{N/2-1}j^m, 
\end{equation}
then the free energy $F^{SO}(\lambda, N)$ may be written explicitly as 
\begin{eqnarray}
\label{chek}
F^{SO}(\lambda,N)&=&-\frac{1}{2}\sum_{m=1}^{\infty}\sum_{j=1}^{N-2}{(}N-1-j)j^m\Bigl{(}\frac{-\lambda}{2\pi (N-1)}\Bigr{)}^m \frac{\zeta(m)}{m}\nonumber \\
&-&\sum_{m=1}^{\infty}\sum_{j=1}^{N/2-1}{(}1-2^{m-1})j^m\Bigl{(}\frac{-\lambda}{2\pi (N-1)}\Bigr{)}^m \frac{\zeta(m)}{m}.
\end{eqnarray}
If we let
 \begin{equation*}F^{SO}(\lambda,N)=F^{O}(\lambda,N)+F^{NO}(\lambda,N),
 \end{equation*}
 where 
\begin{equation}
F^{O}(\lambda,N)=-\frac{1}{2}\sum_{m=1}^{\infty}\sum_{j=1}^{N-2}{(}N-1-j)j^m\Bigl{(}\frac{-\lambda}{2\pi (N-1)}\Bigr{)}^m \frac{\zeta(m)}{m}.
\end{equation}
 Using  the power sum formula
\begin{eqnarray}
\sum_{j\geq 1}^{N-2}j^m &=& \frac{(N-1)^m}{m+1}-\frac{1}{2}(N-1)^m+\frac{1}{m+1}\sum_{g=1}^{[\frac{m}{2}]}\binom{m+1}{2g}B_{2g}(N-1)^{m+1-2g},
\end{eqnarray}
one has  
\begin{eqnarray}
F^{O}(\lambda,N)&=&-\frac{1}{2}\sum_{m=1}^{\infty}\frac{(N-1)^2}{m(m+1)(m+2)}\Bigl{(}\frac{-\lambda}{2\pi }\Bigr{)}^m \zeta(m)\nonumber \\
&-&\frac{1}{2}\sum_{m=1}^{\infty}\sum_{g=1}^{[m/2]}\frac{B_{2g}}{(2g)!}\frac{(m-1)!}{(m+2-2g)}(1-2g)\Bigl{(}\frac{-\lambda}{2\pi (N-1)}\Bigr{)}^m {\zeta(m)},
\end{eqnarray}
and letting $m=2g-2+n$,  we obtain 
\begin{equation}
F^{O}(\lambda,N)=\frac{1}{2}\sum_{\substack{g\ge 1, n>0\\1-g-n<0}} 
 \frac{(2g+n-3)!(2g-1)}{(2g)!(n)!}B_{2g}(N-1)^{2-2g} (\frac{-\lambda}{2\pi })^{2g+n-2}\zeta(2g-2+n),
 \end{equation}
this shows that 
\begin{equation}
F^{O}(\lambda,N)=\frac{1}{2}\sum_{\substack{g\ge 0, n>0\\2-2g-n<0}} 
 \chi^{O}(\mathfrak M_{g,n})(N-1)^{2-2g} (\frac{\lambda}{2\pi })^{2g+n-2}\zeta(2g-2+n),
 \end{equation}
   that is, $F^{SO}(\lambda,N)$, is the generating function for the orbifold Euler characteristic $\chi^{O}(\mathfrak M_{g,n})$ given by equation (\ref{chior}).
   We now move  to compute the last term in equation (\ref{chek}), namely,  
   \begin{equation}
   F^{NO}(\lambda,N)=\sum_{m=1}^{\infty}\sum_{j=1}^{N/2-1}{(}1-2^{m-1})j^m\Bigl{(}\frac{-\lambda}{2\pi (N-1)}\Bigr{)}^m \frac{\zeta(m)}{m},
   \end{equation}
   using the power sum formula 
   \begin{equation}
      \sum_{j\geq 1}^{N/2-1}j^m =\frac{1}{m+1}(\frac{N-1}{2})^m+\frac{1}{m+1}\sum_{g=1}^{[\frac{m}{2}]}(2^{1-2g}-1)\binom{m+1}{2g}B_{2g}(\frac{N-1}{2})^{m+1-2g},
   \end{equation}
   $ F^{NO}(\lambda,N)$ takes the following form 
   \begin{eqnarray}
    F^{NO}(\lambda,N)&=&-\frac{1}{2}\sum_{m=1}^{\infty}\frac{(N-1)}{m(m+1)}\Bigl{(}\frac{-\lambda}{4\pi}\Bigr{)}^m\zeta(m)+\frac{1}{4}\sum_{m=1}^{\infty}\frac{(N-1)}{m(m+1)}\Bigl{(}\frac{-\lambda}{2\pi}\Bigr{)}^m\zeta(m)\nonumber \\&+&\sum_{m=1}^{\infty}\sum_{g\geq1}^{[m/2]}(2^{2g-1}-1)\Bigl{(}\frac{-\lambda}{4\pi}\Bigr{)}^m(N-1)^{1-2g}\frac{B_{2g}}{(2g)!}\frac{(m-1)!}{(m+1-2g)!}\zeta(m)\nonumber \\&-&\frac{1}{2}\sum_{m=1}^{\infty}\sum_{g\geq1}^{[m/2]}(2^{2g-1}-1)\Bigl{(}\frac{-\lambda}{2\pi}\Bigr{)}^m(N-1)^{1-2g}\frac{B_{2g}}{(2g)!}\frac{(m-1)!}{(m+1-2g)!}\zeta(m).
   \end{eqnarray}
Now let $m=2g+n-1$, to obtain 
  %%%%%%%%%%%%%%%%%%%%%%%%%%%%%%%%%%%%
\begin{eqnarray}
\label{nos}
    F^{NO}(\lambda,N)&=& \sum_{\substack{n\geq0, g\geq0\\1-2g-n\le 0}}^{\infty}(2^{2g-1}-1)\frac{(2g+n-2)!}{n!(2g)!}B_{2g}(N-1)^{1-2g}\Bigl{(}\frac{-\lambda}{4\pi}\Bigr{)}^{2g+n-1}\zeta(2g+n-1)\nonumber \\
    &-&\frac{1}{2} \sum_{\substack{n\geq0, g\geq0\\1-2g-n\le 0}}^{\infty}(2^{2g-1}-1)\frac{(2g+n-2)!}{n!(2g)!}B_{2g}(N-1)^{1-2g}\Bigl{(}\frac{-\lambda}{2\pi}\Bigr{)}^{2g+n-1}\zeta(2g+n-1).\nonumber \\
   \end{eqnarray}
   %%%%%%%%%%%%%%%%%%%%%%%%%%%%%%%%%%%
This equation may also be written in terms of the orbifold Euler characteristic of the moduli space of the non-orientable Riemann surfaces,    then the  free energy $ F^{SO}(\lambda,N)$, becomes
  %%%%%%%%%%%%%%%%%%%%%%%%%%%%%%%%%% 
\begin{eqnarray}
\label{extop2}
{F^{SO}}(\lambda,N)&=&\frac{1}{2}\sum_{\substack{g\ge 0, n>0\\2-2g-n<0}} 
 \chi^{O}(\mathfrak M_{g,n})(N-1)^{2-2g} \Bigl{(}\frac{\lambda}{2\pi}\Bigr{)}^{2g+n-2}\zeta(2g+n-2)\nonumber \\ 
&-&\sum_{\substack{g\ge 0,n>0\\ 1-2g-n<0}}
 \chi^{NO}(\mathfrak M_{g,n})
(N-1)^{1-2g} \Bigl{(}\frac{\lambda}{2\pi}\Bigr{)}^{2g+n-1}\zeta(2g+n-1)\nonumber  \\ &+&2\sum_{\substack{g\ge 0,n>0\\ 1-2g-n<0}}
 \chi^{NO}(\mathfrak M_{g,n})
(N-1)^{1-2g} \Bigl{(}\frac{\lambda}{4\pi}\Bigr{)}^{2g+n-1}\zeta(2g+n-1).
\end{eqnarray}
%%%%%%%%%%%%%%%%%%%%%%%%%%%%%%%%%%%%%%%
This shows that the perturbative free energy of Chern-Simons gauge theory with coupling constant $\lambda$, namely, $F^{SO}(\lambda,N)$, generates the virtual  orbifold Euler characteristic  of the moduli space of  orientable and non-orientable Riemann surfaces for all genera $g\geq 0$ with $n$ punctures, i.e,  $\chi^{O}(\mathfrak M_{g,n})$ and $\chi^{NO}(\mathfrak M_{g,n})$ respectively.  This also, proves  our proposition given  in section \ref{sec3}.
%%%%%%%%%%%%%%%%%%%%%%%%%%%%%%%%%%%%%%%%
\section{The Continuum Limit of the Perturbative $SO(N)/Sp(N)$ Chern-Simons Theory}\label{continuum}
%%%%%%%%%%%%%%%%%%%%%%%%%%%%%%%%%%%%%%%%%%%%%%%
 Having identified the Perturbative $SO(N)$ Chern-Simons free energy with the 
deformed Orthogonal Penner model, we  may use the latter to compute the continuum (double scaling) 
limit  of the theory. In  \cite{chair} it was noted   that  in order to obtain  the continuum limit of the perturbative $SU(N)$  Chern-Simons gauge  theory one has to sum   over all boundaries  which  is equivalent to sum over all punctures in  the Penner model, here, we will follow the same procedure to find the continuum limit of the $SO(N)$ Chern-Simons gauge theory. As the sum over all punctures of the deformed $SO(N)$ Penner free energy was obtained in section \ref{sec3}, the continuum limit of the $SO(N)$ Chern-Simons gauge theory is obtained  by defining a new coupling constant\footnote{Here,  we use the same coupling constant that gives the continuum limit of the $SU(N)$ Chern-Simons \cite{das}.} $\nu_n$ given  by  $\nu_n=\frac{2\pi(2N)}{\lambda}(\frac{\lambda}{2\pi}-n)$ in the expressions for the free energy.  Using equation  (\ref{z}) for $g=0$,  one has
%%%%%%%%%%%%%%%%%%%%%%%%%%%%%%
\begin{eqnarray}
\mathbf{F}^{NO}_{0}(\nu,N)&\cong&\sum_{n\in \mathbf Z^*}(-1)^{n+1}\Bigl{[}\frac{\nu_{n}}{4}\log\nu_{n}+\frac{\nu_{n}}{4}
\log\frac{\lambda}{4\pi nN}+\frac{i\pi\nu_{n}}{2}\Bigr{]},
\end{eqnarray}
%%%%%%%%%%%%%%%%%%%%%%%%
where $\mathbf Z^*$  is the set of all positive and  negative  integers. Similarly  for  higher genus $g\geq1$ equation (\ref{lambdasum}), gives
%%%%%%%%%%%%%%%%%%%%%%%%
\begin{equation}
\mathbf{F}^{NO}_{g\geq1}(\lambda,N)=\sum_{g\geq1}^{\infty}\sum_{ {n\in \mathbf Z^*}
}\chi^{NO}(\mathfrak M_{g,0})(-1)^{n}\nu_{n}^{1-2g}.
\end{equation}
%%%%%%%%%%%%%%%%%%%%%%%%%%%%%%%%%%%%
The $SO(N)$ Chern-Simons coupling constant $\lambda$ is related to the level of Kac-Moody algebra $k$, by $\lambda=\frac{2\pi(N-1)}{(k+N-2)}$, this shows that $\lambda$ has a fundamental domain between $0$ and $2\pi$. Therefore, the natural critical double scaling limit would be
%%%%%%%%%%%%%%%%%%%%%%%%%%%%%%%%%%% 
\begin{equation*}
\lambda\rightarrow 2\pi \hspace{.5 in} \nu_1=\text{finite}.
\end{equation*}
%%%%%%%%%%%%%%%%%%%%%%%%%%%%%%%%%%%
Using this limit and keeping only the non analytic terms in the total free energy,  one has 
%%%%%%%%%%%%%%%%%%%%%%%%%%%%%%%%%
\begin{eqnarray}
F^{NO}(\nu_{1})&=&\frac{\nu_1}{4}\log{\nu_1}
-\sum_{g\geq1}^{\infty}
\chi^{NO}(\mathfrak M_{g,0})\nu_{1}^{1-2g}.
\end{eqnarray} 
%%%%%%%%%%%%%%%%%%%%%%%%%%%%%%%%%%%%%%55
Therefore, including  the orientable contributions to the $SO(N)$ Chern-Simons gauge theory \cite{chair, das}, then  the full continuum limit of the perturbative $SO(N)$ Chern-Simons reads
%%%%%%%%%%%%%%%%%%%%%%%%%%%%%%%%%%%%%%%55
\begin{eqnarray}
\label{man}
\mathbf{F}^{SO}(\nu_{1})&\cong&\frac{{\nu_1}^2}{4}\log{\nu_1}-\frac{1}{24}\log\nu_1+\frac{\nu_1}{4}\log{\nu_1}-\frac{1}{24\nu_1}\nonumber \\
&+&\frac{1}{2}\Big{(}\sum_{g\geq2}^{\infty}
\chi^{O}(\mathfrak M_{g,0})\nu_{1}^{2-2g}-2\sum_{g\geq2}^{\infty}
\chi^{NO}(\mathfrak M_{g,0})\nu_{1}^{1-2g}\Big{)}.
\end{eqnarray}
It is interesting to note  that this expression  is  equivalent to the continuum limit of the  $SO(N)$ Penner model\footnote{This  also appears in the $SU(N)$ Chern-Simons gauge theory\cite{chair}.} when $\nu_1$ is replaced with $-\mu$  \cite{madalabeeh}.
Simalrly, the continuum limit of the perturbative $Sp(N)$ Chern-Simons gauge theory\textbf{} is
\begin{eqnarray}
\label{man1}
\mathbf{F}^{Sp}(\nu_{1})&\cong&\frac{{\nu_1}^2}{4}\log{\nu_1}-\frac{1}{24}\log\nu_1-\frac{\nu_1}{4}\log{\nu_1}+\frac{1}{24\nu_1}\nonumber \\
&+&\frac{1}{2}\Big{(}\sum_{g\geq2}^{\infty}
\chi^{O}(\mathfrak M_{g,0})\nu_{1}^{2-2g}+2\sum_{g\geq2}^{\infty}
\chi^{NO}(\mathfrak M_{g,0})\nu_{1}^{1-2g}\Big{)}.
\end{eqnarray}
In the next section we will  compute  $\log\text{vol}(G)$ for $G= SO(2N+1)$ and $G=Sp(2N-1) $ and show that the continuum limit of the perturbative  $SO(N)/Sp(N)$ Chern-Simons gauge theories are reproduced. 
%%%%%%%%%%%%%%%%%%%%%%%%
\section{The Gauge Group Volumes}\label{nonpert}
%%%%%%%%%%%%%%%%%%%
 Ooguri and Vafa \cite{oguri}  related $\log \text{vol}(G)$ for $G=SO(2N+1), \text{and}\ G=Sp(2N-1)$ to the virtual Euler characteristic without punctures on the moduli space of orientable and non-orientable Riemann surfaces for genus $g\geq 2$. Here, we will give an alternative derivation for the asymptotic expansions of $\log (\text{vol}(SO(2N+1)))$ and $\log( \text{vol}(Sp(2N-1)))$ for all genera $g\geq0$. It will be shown that the latter is  related to the continuum limit of the perturbative $SO(N)/Sp(N)$ Chern-Simons gauge theories given in the last section. We  will also show that the volume of the gauge groups $SO(2N)$, $Sp(2N)$ are equivalent  to the continuum limit of the $SO(N)/Sp(N)$ Penner models \cite{madalabeeh}.  
The expression for $\text{ vol}(SO(2N+1))$ \cite{oguri}, is given by \footnote{This formula may also be found in the work of  L. K. Hua  in his classic text book \cite{china}.}
%%%%%%%%%%%%%%%%%%%%%%%%%%%%%%% 
\begin{equation}
\label{sam}
\text{vol}(SO(2N+1))=\frac{2^{N+1}(2\pi)^{N^2+N-\frac{1}{4}}}{(2N-1)!(2N-3)!\cdots   3!1!}.
\end{equation} 
%%%%%%%%%%%%%%%%%%%%%%%%%%%%%%
 This  may be written in terms of the Barnes function $G_2(z)$  \cite{Adamchik, oguri},  defined by 
%%%%%%%%%%%%%%%%%%%%%%%%%%%%%
\begin{equation*}
G_2(N+1)=\prod_{k=1}^{N-1}(N-k)!,
\end{equation*} 
%%%%%%%%%%%%%%%%%%%%%%%%%%%%%%%%%%%%%%
to give 
%%%%%%%%%%%%%%%5
\begin{equation}
\text{vol}(SO(2N+1))=\frac{2^{N+1}(2\pi)^{N^2+N-\frac{1}{4}}}{G_2(2N+1)}\prod_{k=1}^{N-1}(2N-2k)!.
\end{equation}
%%%%%%%%%%%%%%%%%%%%%%%%%%%%%%%%%%%%%%%%
Using the following identity 
%%%%%%%%%%%%%%%%%%%%%%%%%%%%%%%%%%%%%%%%%
\begin{eqnarray}
\label{amman}
\log\prod_{k=1}^{N-1}(2N-2k)!&=&\frac{1}{2}\log G_2(2N+1)-\log(2N-1)!!,
\end{eqnarray}
%%%%%%%%%%%%%%%5
where $(2N-1)!!=\frac{2^N}{\sqrt\pi}\Gamma(N+\frac{1}{2}),$
then it is easy to show that
%%%%%%%%%%%%%%%%%%%%%%%%%%%%
\begin{equation}
\log(\text{vol}(SO(2N+1)))\cong-\frac{1}{2}\log G_2(2N+1)-\frac{1}{2}\log\Gamma(N+\frac{1}{2}),
\end{equation}
%%%%%%%%%%%%%%%%%%%%%%%%%%%%
 note that we have discarded terms which do not give rise to singularities.  In the large $N$ limit we may expand $\log\Gamma(N+\frac{1}{2})$, \cite{bernouli} as follows
%%%%%%%%%%%%%%%%%%%%%%%%%
\begin{equation}
\log\Gamma(N+\frac{1}{2})\cong N\log N+\sum_{g=1}^\infty\frac{B_{2g}(\frac{1}{2})}{2g(2g-1)}(N)^{1-2g},
\end{equation}
%%%%%%%%%%%%%%%%%%%%
 where  $ B_{2g}(\frac{1}{2})=(2^{1-2g}-1)B_{2g}$, therefore, the above equation may be rewritten as 
%%%%%%%%%%%%%%%%%%%%%%%%%%%%
\begin{equation}
\label{gamma}
\log\Gamma(N+\frac{1}{2})\cong N\log 2N-\sum_{g=1}^\infty\frac{B_{2g}(2^{2g-1}-1)}{2g(2g-1)}(2N)^{1-2g}.
\end{equation}
%%%%%%%%%%%%%%%%%%%%%%%%%%%%
On the other hand, the  large $N$ expansion for $\log G_2(2N+1)$ \cite {oguri,Adamchik}, is 
%%%%%%%%%%%%%%%%%%%%%%%%%%%%%%
\begin{equation}
\log G_2(2N+1)\cong\frac{(2N)^2}{2}\log2N-\frac{1}{12}\log2N+\sum_{g=2}^\infty\frac{B_{2g}}{2g(2g-2)}(2N)^{2-2g}.
\end{equation}
%%%%%%%%%%%%%%%%%%%%%%%%%%%%%
Finally, the  large $N$  expansion for $\log (\text{vol}(SO(2N+1)))$ to all genera reads 
%%%%%%%%%%%%%%%%%%%%%%%%%%%%%%%%%%
\begin{eqnarray}
\log (\text{vol}(SO(2N+1)))&\cong &-\frac{(2N)^2}{4}\log2N+\frac{1}{24}\log2N-\frac{(2N)}{4}\log 2N+\frac{1}{24}(2N)^{-1}\nonumber \\&-&\sum_{g=2}^\infty\frac{B_{2g}}{4g(2g-2)}(2N)^{2-2g}+\sum_{g=2}^\infty\frac{B_{2g}(2^{2g-1}-1)}{4g(2g-1)}(2N)^{1-2g},
\end{eqnarray}
%%%%%%%%%%%%%%%%%%%%%%%%%%%%%%%%%%%%%
this expression may be written using the orbifold Euler characteristic as follows
%%%%%%%%%%%%%%%%%%%%%%%%%%%%%%%%%
\begin{eqnarray}
\label{torus1}
-\log (\text{vol}(SO(2N+1)))&\cong &\frac{(2N)^2}{4}\log2N-\frac{1}{24}\log2N+\frac{(2N)}{4}\log 2N-\frac{1}{24}(2N)^{-1}\nonumber \\&+&\frac{1}{2}\Bigl{(}\sum_{g=2}^\infty\frac{\chi^O(\mathfrak M_{g,0})}{(2N)^{2g-2}}-2\sum_{g=2}^\infty\frac{\chi^{NO}(\mathfrak M_{g,0})}{(2N)^{2g-1}}\Bigr{)}.
\end{eqnarray}
%%%%%%%%%%%%%%%%%%%%%%%%%%%%%%%%%
  Similarly, the expression for $-\log (\text{vol}(Sp(2N-1)))$, is
  %%%%%%%%%%%%%%%%%%%%%%%%%%%%%%%%%
  \begin{eqnarray}
  \label{torus2}
-\log(\text{vol}(Sp(2N-1)))&\cong &\frac{(2N)^2}{4}\log2N-\frac{1}{24}\log(2N)-\frac{(2N)}{4}\log 2N+\frac{1}{24}(2N)^{-1}\nonumber \\&+&\frac{1}{2}\Bigl{(}\sum_{g=2}^\infty\frac{\chi^O(\mathfrak M_{g,0})}{(2N)^{2g-2}}+2\sum_{g=2}^\infty\frac{\chi^{NO}(\mathfrak M_{g,0})}{(2N)^{2g-1}}\Bigr{)}.
\end{eqnarray}
%%%%%%%%%%%%%%%%%%%%%%%%%%%%%%
In obtaining the above equation we used the following relation \cite{oguri},
%%%%%%%%%%%%%%%%%%%%%%%%%%
  \begin{eqnarray}  
  \log (\text{vol}(SO(2N+1)))+ \log(\text{vol}(Sp(2N-1)))\cong-\frac{(2N)^2}{2}+\frac{1}{12}\log(2N)\nonumber \\-\sum_{g=2}^\infty\frac{B_{2g}}{(2g)(2g-2)(2N)^{2g-2}}.
 \end{eqnarray} 
 %%%%%%%%%%%%%%%%%%%%%%%%%%%%%%%%%%%%%%%%%%
Therefore, comparing  the  terms given by equations (\ref{torus1}) and (\ref{torus2}) with those of the continuum limits of the perturbative $SO(N)/Sp(N)$ Chern-Simons gauge theories given by equations (\ref{man}) and (\ref{man1}) of the last section, we see that  $-\log(\text{vol}(SO(2N+1)))$
is equivalent to the continuum limit of the perturbative  $SO(N)$ Chern-Simons theory, when $2N$   is replaced  by $\nu_1$. We also have  the equivalence between  $-\log(\text{vol}(Sp(2N-1)))$ and the continuum limit of the perturbative  $Sp(N)$ Chern-Simons gauge  theory. This shows that the coefficients in the continuum limit of $SO/Sp$ Chern-Simons gauge theories are exactly those of the nonperturbative contributions given by equations (\ref{torus1}) and (\ref{torus2}), respectively, up to regular terms.  
Similar results for higher genus contributions to the  $\log (\text{vol}(SO(2N+1)))$  may be found in \cite{oguri}, where the authors used  different approach to obtain the expression for $\log(\text{vol}(SO(2N+1)))$. However,  a factor of $+2$ should be inserted  in front of the expression of the Euler characteristic of the moduli space of genus $g$ with a single cross cap instead of minus one, see equation (4.31) in \cite{oguri} and equation (\ref{gamma}) above.
Ogouri and Vafa \cite{oguri}, related the volume of $U(N)$ to the orbifold Euler characteristic $\chi( \mathfrak{M}_{g,0})$ of the moduli space of genus $g$ Riemann surfaces as 
\begin{equation}
-\log(\text{vol}(U(N)))\cong \frac{N^2}{2}\log N -\frac{1}{12}\log N +\sum_{g\geq 2}\frac{\chi(\mathfrak{M}_{g,0})}{N^{2g-2}}.
\end{equation}
Note that, this equation is equivalent to the continuum limit of the Penner model \cite{distler,chairold}, provided one sets $N=\mu$. Here, it will be shown that the volume of the gauge groups $SO(2N)$, $Sp(2N)$ are equivalent  to the continuum limit of the $SO(N)$ and $Sp(N)$ Penner models respectively \cite{madalabeeh}, the volume of the gauge groups $SO(2N)$, $Sp(2N)$ are given by \cite{oguri,china}
\begin{eqnarray}
\label{mas}
\text{vol}(SO(2N))&=&\frac{\sqrt 2(2\pi)^{N^{2}}}{(2N-3)!(2N-5)!\cdot\cdot\cdot3!1!(N-1)!},\nonumber \\
\text{vol}(Sp(2N))&=&\frac{ 2^{-N}(2\pi)^{N^{2}+N}}{(2N-1)!(2N-3)!\cdot\cdot\cdot3!1!},
\end{eqnarray}
then it is not difficult to show that 
\begin{equation}
\label{amman1}
\log(\text{vol}(SO(2N)))-\log(\text{vol}(Sp(2N)))\cong \log\Gamma(N+\frac{1}{2}),
\end{equation}
and 
\begin{equation}
\log(\text{vol}(SO(2N)))+\log(\text{vol}(Sp(2N)))\cong -\log G_2(2N+1),
\end{equation}
as  a result  one has 
\begin{eqnarray}
\label{mmmy}
-\log(\text{vol}(SO(2N)))&\cong& \frac{1}{2}\log G_2(2N+1)-\frac{1}{2}\log\Gamma(N+1/2)\nonumber \\&\cong&\frac{1}{4}(2N)^2\log2N-\frac{1}{24}\log2N-\frac{(2N)}{4}\log2N+\frac{1}{24(2N)} \nonumber \\&+& \frac{1}{2}\sum_{g\geq2}\Bigl{(}\frac{\chi^{O}(\mathfrak{M}_{g,0})}{(2N)^{2g-2}}+2\frac{\chi^{NO}(\mathfrak{M}_{g,0})}{(2N)^{2g-1}}\Bigr{)}.
\end{eqnarray}
Similarly, 
\begin{eqnarray}
\label{ttty}
-\log(\text{vol}(Sp(2N)))&\cong& \frac{1}{2}\log G_2(2N+1)+\frac{1}{2}\log\Gamma(N+\frac{1}{2})\nonumber \\&\cong&\frac{1}{4}(2N)^2\log2N-\frac{1}{24}\log2N+\frac{(2N)}{4}\log2N-\frac{1}{24(2N)} \nonumber \\&+& \frac{1}{2}\sum_{g\geq2}\Bigl{(}\frac{\chi^{O}(\mathfrak{M}_{g,0})}{(2N)^{2g-2}}-2\frac{\chi^{NO}(\mathfrak{M}_{g,0})}{(2N)^{2g-1}}\Bigr{)}.
\end{eqnarray}
If we let $2N=\mu$, then $-\log(\text{vol}(SO(2N)))$, $\log(\text{vol}(Sp(2N)))$ are  the continuum limit of the $SO(2N)$ and $Sp(N)$ Penner models respectively  \cite{madalabeeh}, see equation (\ref{finalsym}) in this paper.
%%%%%%%%%%%%%%%%%%%%%%%%%%
\section{Discussion} 
In this paper we have related the $SO/Sp$ Chern-Simons  gauge theories at large $N$ to the $SO/Sp$ Penner models through   deformed $SO/Sp$ Penner generation functions by construction. This construction is then proved to be correct using a sector in the perturbative $SO(N)$ Chern-Simons free energy   $F^{SO}(\lambda,N)$, with coupling constant $\lambda$, also it was shown that the latter generates the virtual orbifold Euler  characteristic of the moduli space of orientable and non-orientable Riemann surfaces of genus $g$ with $n$ punctures. However, there is  no restriction on the number of punctures unlike the $SO(N)$ Chern Simons gauge theory  puts restrictions on  the number of punctures. This connection enables us to think of the perturbative $SO(N)/Sp(N)$ Chern-Simons gauge theory as two deformed $SO(N)$  Penner models of opposite coupling constants summed over all  instantons. On the other hand, when summing over all punctures we end up with the perturbative $SO(N)$  Chern-Simons free energy written in terms of the orbifold Euler characteristic without punctures, since we used  the free energy of  Penner model which is known  when summed  over all punctures  one obtains a  free energy that contains the virtual Euler characteristic without punctures. Also in this paper we clarified the disappearance of the contribution from constant maps in the   non-orientable part of the perturbative $SO(N)$ Chern-Simons free energy, through the Penner model.  Both the free energy and the continuum limit in the $SO/Sp$ Chern-Simons gauge theories were obtained using the $SO/Sp$ Penner models. This is an extension to the work of Das and Gomez \cite{das}. Also, it was shown that the  asymptotic expansions of $\log(\text{vol}(G))$ where $G=SO(2N+1)$, $G=Sp(2N-1)$ are equivalent to the continuum limit of the perturbative $SO/Sp$ Chern-Simons  gauge theories respectively. However, when $G=SO(2N)$, $G=Sp(2N)$, the asymptotic expansions of the logarithm of the gauge groups are equivalent to continuum limit of the $SO/Sp$ Penner models. Our computation of the  asymptotic  expansion for $\log(\text{vol}(SO(2N+1))$ is  different from that of Ooguri and Vafa \cite{oguri}, and the result  contains all genera. Comparing the two   free energies  of  $SO(N)/Sp(N)$ Penner models and  $SO(N)/Sp(N)$ Chern-Simons gauge theories it is clear that  both  have the same structure namely 
\begin{equation}
F^{SO/Sp}=F^{O}\mp F^{NO},
\end{equation}
where the first term corresponds   to half of the $SU(N)$ contribution, while, the second term is the non-orientable contribution.
It was shown that  the universal behavior of the topological partition function of the topological field theory  (the Kodaira-Spencer theory) of the Calabi-Yau threefold near a conifold singularity   was given by the free energy of the  Penner model in the continuum limit \cite{goshal}. Also it was shown that this topological partition function reduces to  Chern-Simons theory on $S^3$ \cite{stony}. From  this work, it is clear that  the behavior of the continuum limit of  both $SO/SP$ Chern-Simons  and $SO/Sp$ Penner  free energies are equivalent, Therefore,  the topological partition function on the quotient of the resolved conifold  by involution is  given by the $SO/Sp$ Penner free energy. This connection, however,  needs both  physical and mathematical explanation.  
\\ \ Next we will make some remarks on the relations between different gauge group volumes appeared in this paper, and the connection of these  volumes with the generating function  $\log\prod_{p\ \text{odd}}^{2N-1}(1+pt)^{\frac{1}{2}}$ for $\chi^{NO}{(\mathfrak{M}_{g,n}})$ identified in \cite{madalabeeh}. From  the  formulae  for  $\text{vol}(SO(2N+1))$ and $\text{vol}(Sp(2N))$, given by equations (\ref{sam}), and (\ref{mas}), it follows that their asymptotic expansions should be the same, see equations (\ref{torus1}) and (\ref{ttty}). Similarly $-\log(\text{vol}(Sp(2N-1)))$ and $-\log(\text{vol}(SO(2N)))$, have the same asymptotic expansions see equations (\ref{torus2}) and (\ref{mmmy}). This equivalence should follow from the volume of the gauge groups $G=Sp(2N-1)$ and  $G=SO(2N)$, to that end let $N\rightarrow N-\frac{1}{2}$,  in the expression for $\text{vol}(Sp(2N))$, to give 
\begin{equation}
\text{vol}(Sp(2N-1))\cong \prod_{k=1}^{N-1}\frac{1}{(2N-2k)!}.
\end{equation}
On the other hand, one can show that 
\begin{equation}
\text{vol}(SO(2N))\cong \prod_{k=1}^{N-1}\frac{(2N-2k)!}{G_2(2N+1)}\Gamma(N+\frac{1}{2}), 
\end{equation}
and from equation (\ref{amman}), it follows that
\begin{equation}
\Bigl{(}\prod_{k=1}^{N-1}{(2N-2k)!}\Bigr{)}^2\cong\frac{G_2(2N+1)}{\Gamma(N+\frac{1}{2})}, 
\end{equation} 
hence  the equivalence of the $\text{vol}(SO(2N))$ and $\text{vol}(Sp(2N-1))$. From equation (\ref{amman1}) it is seen that 
\begin{equation}
\label{80}
\mp\frac{1}{2}\log\Bigl{(}\frac{\text{vol}(SO(2N))}{\text{vol}(Sp(2N))}\Bigr{)}\cong \mp\frac{(2N)}{4}\log(2N)\pm \sum_{g\geq 1}\chi^{NO}(\mathfrak M_{g,0})\frac{1}{(2N)^{2g-1}},
\end{equation}
that is, $\mp\frac{1}{2}\log\Bigl{(}\frac{\text{vol}(SO(2N))}{\text{vol}(Sp(2N))}\Bigr{)}$  generates the non-orientable contributions to the free energy of the $SO/Sp$ Penner models  in the continuum limit. On the other hand, it is well known from Penner model that differentiating the free energy in the continuum limits $n$ times with respect to the continuum variable $\mu$ brings back the punctures to the Riemann surface \cite{distler, chairold}. Hence , differentiating equation (\ref{80}) $n$ times with respect to $2N$ will produce a generating function for the orbifold Euler characteristic of the moduli space of non-orientable Riemann surfaces of   genus $g$ with $n$ punctures.
%%%%%%%%%%%%%%%%%
\newpage
\bibliographystyle{phaip}

\end{document}